\documentclass{article}
\usepackage{amssymb}
\usepackage{amsfonts}
\usepackage{amsmath}
\usepackage{graphicx}% Include figure files
\begin{document}

\title{Hamiltonian Dynamics of the Protein Chain and Normal Modes of
$\alpha$-Helix and $\beta$-Sheet}
\author{Hon-Wai Leong$^{1}$, Lock-Yue Chew$^{1}$ and Kerson Huang$^{2}$ \\
%EndAName
\\
$^{1}$ Division of Physics \& Applied Physics, \\
School of Physical \& Mathematical Sciences, \\
Nanyang Technological University,\\
SPMS-04-01, 21 Nanyang Link, Singapore 637371\\
$^{2}$Physics Department, Massachusetts Institute of Technology,\\
Cambridge MA, USA 02139}
\maketitle

\begin{abstract}
We use the torsional angles of the protein chain as generalized coordinates
in the canonical formalism,\ derive canonical equations of motion, and
investigate the coordinate dependence of the kinetic energy expressed in
terms of the canonical momenta. We use the formalism to compute the
normal-frequency distributions of the $\alpha $-helix and the $\beta $%
-sheet, under the assumption that they are stabilized purely
through hydrogen bonding. Comparison of their free energies show
the existence of a phase transition between the $\alpha $-helix
and the $\beta $-sheet at a critical temperature.
\end{abstract}

\section{Introduction and results}

The purpose of this work is to describe the backbone chain of a
protein molecule in terms of dynamically independent variables,
which are the torsional angles between successive units of the
chain. These angles determined the average conformation of the
chain, and local vibrations of chemical bonds only contribute to
small fluctuations about the average. By ignoring these
fluctuations, we gain a better overview of the motion of the
chain, in particular its folding.

We use the torsional angles as generalized coordinates in the
canonical formalism, with concomitant canonical momenta. The
kinetic energy then becomes a function of the coordinates, when
expressed in term of the canonical momenta. That is, masses are
replaced by a generalized mass matrix, which is a function of the
coordinates. There arises an effective potential, which is
discussed in detailed later. By studying this mass matrix
numerically, we find that the effective potential is approximately
constant for almost all conformations of the chain. This result is
significant for practical applications, particularly for the CSAW
(conditioned self-avoiding walk) model \cite{Huang,Huang02}, where
such an approach was first used.

Using the canonical formalism, we formulate the eigenvalue problem that
describes the normal modes of the system with respect to an equilibrium
conformation. Actual computations are carried out for a pure $\alpha $-helix
and a pure $\beta $-sheet, to obtain distribution functions of the normal
frequencies. These pure structures are hypothetical, of course, for in a
real situation they are embedded inside a larger protein. However, we can
learn something useful from these examples.

First of all, in our model we assume that the $\alpha $-helix and
the $\beta $-sheet are stabilized purely through hydrogen bonding.
The positive-definiteness of normal frequencies indicates that
these structures can maintain mechanical stability from hydrogen
bonding. This leads one to expect that in the unfolded  protein
chain, which is subject to random forces from the solution, these
secondary structures may still have transient existence.

The normal-frequency distribution function for the $\alpha $-helix
exhibits a number of peaks. By examining the corresponding
eigenvectors, we can associate them with types of distortion,
namely stretching, twisting and bending. By superposing the
distributions of several $\alpha $-helices, we can construct an
approximate normal-frequency distribution of an all-$\alpha $
protein, such as myoglobin.

We can obtain the free energy of a structure near equilibrium by
treating the system as a collection of harmonic oscillators with
the calculated normal frequencies. As such an exercise, we compute
the free energy of a pure $\alpha $-helix and that of a pure
$\beta$-sheet, and plot the results as functions of temperature.
We find that the two curves intersect, indicating a phase
transition occurring at that temperature. Such a model is of
course too crude to have quantitative significance, for real
secondary structures are embedded in a larger protein, and
interactions not taken into account here may be important.
However, in view of the importance of the
subject, in particular its possible relevance to the prion transition \cite%
{Prusiner}, any exploratory calculation in this direction might not be
totally meaningless.

\section{Modelling the protein chain}

\begin{center}
\begin{figure}
    \includegraphics[scale=0.6]{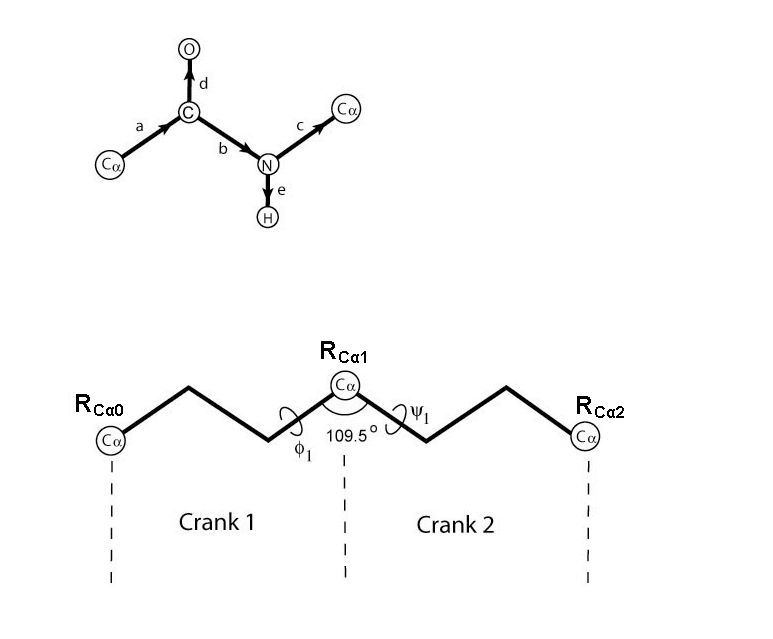}
    \caption{Upper panel shows the
``crank" that connects the center of one residue to the next. The
vectors \textbf{a, b, c, d, e} represent chemical bonds, which all
lie in the same plane. (Side chains have been omitted for
clarity). Lower panel shows how the cranks are
connected to form the backbone of the protein. The vector position of a C$_{%
\protect\alpha }$ is denoted by \textbf{R$_{\bf{C\alpha
\it{i}}}$}. The angle between cranks being fixed$,$ the relative
orientation of successive cranks is specified by two torsional
angles $\protect\phi$ and $\protect\psi $. The conformation of the
backbone chain is completely specified by a set of torsional
angles. Data for bond lengths and angles are given in Table
\ref{tab1}.}\label{fig1}
\end{figure}
\end{center}

The protein chain consists of a sequence of amino acids chosen from a pool
of 20. These amino acids all center about a carbon atom called the C$%
_{\alpha }$, and differ from one another only in the side chains
connected to the C$_{\alpha }$. When the amino acids are joined
into a chain, they become interlocked ``residues". From a
dynamical point of view, the independent units of the chain are
``cranks" made up of coplanar chemical bonds, which connect one
C$_{\alpha }$ to the next, as shown in Fig. \ref{fig1}. The bond
lengths and bond angles in a crank are given in Table \ref{tab1}
\cite{wwpdb}.

\begin{table}[tbp]
\caption{Bond lengths and bond angles. Data obtained from Protein
Data Bank web site (http://www.pdb.org)\cite{wwpdb}.}
\label{tab1}%
\begin{center}
\begin{tabular}{|c|c||c|c|}
\hline & Bond Length (\AA) &  & Bond Angle $(^{\circ })$ \\ \hline
C$_{\alpha }$-C & 1.525 & $\angle $C$_{\alpha }$CN & 116.2 \\
C-N & 1.329 & $\angle $CNC$_{\alpha }$ & 121.7 \\
N-C$_{\alpha }$ & 1.458 & $\angle $NC$_{\alpha }$C & 109.5 \\
C-O & 1.231 & $\angle $C$_{\alpha }$CO & 120.8 \\
N-H & 1.000 & $\angle $C$_{\alpha }$NH & 114.0 \\ \hline
\end{tabular}%
\end{center}
\end{table}

The backbone of the protein chain is thus a sequence of cranks.
The angle between two adjacent cranks is fixed at the tetrahedral
angle $\cos ^{-1}(-1/3)\approx 109.5^{\circ}.$ Thus, the
orientation of one crank with respect to its predecessor is
specified by two torsional angles \{$\phi ,\psi \},$ as
illustrated in Fig. \ref{fig1}.

The conformation of the backbone of the protein is completely
specified by a set of torsional angles $\{\phi _{1},\psi _{1};\phi
_{2},\psi _{2};\cdots \}. $ In this study, we only consider these
torsional degrees of freedom, ignoring the small high-frequency
vibrations within the cranks. Such a description has been used in
the CSAW model (conditioned self-avoiding walk) of protein
folding. \cite{Huang, Huang02}

\section{Canonical formalism}

Consider a chain of $n$ cranks. Let
\begin{eqnarray}
\mathbf{R}_{si} &=&\text{ position vector of atom }s\text{ on crank }i
\notag \\
i &=&1,\ldots ,n  \notag \\
s &=&\text{ C}_{\alpha }\text{, C, N, O, H}
\end{eqnarray}%
We also write this as $\mathbf{R}_{i}$, an array of the position
vectors of elements labelled by $ s$.

In this study we do not take the side chains into account, so the
closest real protein to our model is polyglycine. Formally, it is
straightforward to generalize the model.

Let the set of torsional angles be $\{\phi _{i},\psi
_{i}\}(i=1,\ldots ,n-1).$ Assuming the cranks to be perfectly
rigid places constraints on the vector positions. The constraints
are solved by using the torsional angles as generalized
coordinates, which we denote by the notation
\begin{equation}
q_{ik}\text{ \ }\left( i=1,\ldots ,n-1;\text{ }k=1,2\right).
\end{equation}%
Thus, for example,%
\begin{eqnarray}
q_{11} = \phi _{1}&&q_{12} = \psi _{1}  \notag \\
q_{21} = \phi _{2}&&q_{22} = \psi _{2} \text{\ \ \ \ etc.}
\end{eqnarray}%
We also use the notation $q_{\alpha }$, where $\alpha =\{i,k\}.$ We are to
regard $\mathbf{R}_{i}$ as functions of $\{q_{\alpha }\}$.

The velocity is given by

\begin{equation}
\mathbf{\dot{R}}_{i}=\sum_{i=1}^{n-1}\sum_{k=1}^{2}\frac{\partial \mathbf{R}%
_{i}}{\partial q_{ik}}\dot{q}_{ik}=\sum_{\alpha }\frac{\partial \mathbf{R}%
_{i}}{\partial q_{\alpha }}\dot{q}_{\alpha }.
\end{equation}%
The total kinetic energy is%
\begin{equation}
K\left( q,\dot{q}\right) =\frac{1}{2}\sum_{i=1}^{n-1}m\mathbf{\dot{R}}%
_{i}^{2}=\frac{1}{2}\sum_{\alpha ,\beta }\dot{q}_{\alpha }\left(
\sum_{i=1}^{n-1}m\frac{\partial \mathbf{R}_{i}}{\partial q_{\alpha }}\cdot
\frac{\partial \mathbf{R}_{i}}{\partial q_{\beta }}\right) \dot{q}_{\beta }=%
\frac{1}{2}\dot{q}^{T}M\dot{q}.
\end{equation}%
where the mass matrix is given by:%
\begin{equation}
M_{\alpha \beta }=\sum_{i=1}^{n-1}m\frac{\partial
\mathbf{R}_{i}}{\partial q_{\alpha }}\cdot \frac{\partial
\mathbf{R}_{i}}{\partial q_{\beta }}.
\end{equation}%
This is a symmetric matrix, with $M^{T}=M$. We use the shorthand%
\begin{equation}
m\mathbf{\dot{R}}_{i}^{2}=\sum\limits_{s}m_{s}\mathbf{\dot{R}}_{si}^{2}.
\end{equation}%
where $m_{s}$ denotes the mass of atom $s$ on the crank.

\bigskip The Lagrangian of the backbone chain is given by
\begin{eqnarray}
L(q,\dot{q}) &=&K(\dot{q},q)-U(q)  \notag \\
&=&\frac{1}{2}\dot{q}^{T}M\dot{q}-U(q).  \label{eqLag}
\end{eqnarray}%
Hence, the canonical momentum is
\begin{equation}
p=\frac{\partial L}{\partial \dot{q}}=M\dot{q}.  \label{eq10}
\end{equation}%
and the generalized force is
\begin{equation}
\frac{\partial L}{\partial q}=\frac{1}{2}\dot{q}^{T}\frac{\partial M}{%
\partial q}\dot{q}-\frac{\partial U}{\partial q}.  \label{eq10a}
\end{equation}%
\newline
The Lagrange equation of motion%
\begin{equation}
\frac{d}{dt}\left( \frac{\partial L}{\partial \dot{q}}\right) =\frac{%
\partial L}{\partial q}
\end{equation}
leads to
\begin{equation}
M\ddot{q}=-\frac{1}{2}\dot{q}^{T}\frac{\partial M}{\partial q}\dot{q}-\frac{%
\partial U}{\partial q}.  \label{eqLagEqM}
\end{equation}

\bigskip The Hamiltonian is
\begin{eqnarray}
H(p,q) &=&K(p,q)+U(q)  \notag \\
&=&\frac{1}{2}\dot{q}^{T}M\dot{q}+U(q).  \label{eqHam}
\end{eqnarray}%
From (\ref{eq10}) we have $\dot{q}=M^{-1}p$ and $\dot{q}%
^{T}=p^{T}(M^{-1})^{T}$. Therefore,
\begin{equation}
H(p,q)=\frac{1}{2}p^{T}M^{-1}p+U(q).
\end{equation}%
\newline
The canonical equations of motion%
\begin{equation}
\dot{p}=-\frac{\partial H}{\partial q}, \text{\ }\dot{q}=\frac{\partial H}{%
\partial p}
\end{equation}%
take the forms%
\begin{eqnarray}
\dot{p} &=&-\frac{1}{2}p^{T}\frac{\partial M^{-1}}{\partial q}p-\frac{%
\partial U}{\partial q},  \notag \\
\dot{q} &=&M^{-1}p.
\end{eqnarray}%
\newline
\vspace{-15pt}These are, of course, the same as the Lagrangian
equation of motion (\ref{eqLagEqM}).\\

\section{The effective potential}

The partition function of the system is, up to a constant scale factor,
given by%
\begin{equation}
Z=\int dq\int dp \ e^{-\beta K(p,q)}e^{-\beta U(q)},
\end{equation}%
where $\beta =\left( k_{B}T\right) ^{-1}$ is the inverse temperature. The $p$%
-integration is Gaussian, and can be immediately carried out, and the result
generally depends on $q$. This gives rise to an effective potential $V_{\text{%
eff}}(q),$ which is defined through the relation%
\begin{equation}
e^{-\beta V_{\text{eff}}(q)}\equiv
\left(\frac{\beta}{2\pi}\right)^{n-1}\int dp\ e^{-\beta K(p,q)}.
\end{equation}%
\newline
Thus%
\begin{eqnarray}
Z &=&\int dq\ \rho _{\text{con}}(q),  \notag \\
\rho _{\text{con}}(q) &\equiv &e^{-\beta \left(
U+V_{\text{eff}}\right) },
\end{eqnarray}%
where $\rho _{\text{con}}$ is the configurational probability
density. That is, $\rho _{\text{con}}dq$ is the relative
probability of finding the system in $dq,$ regardless of momentum
$p$. If the kinetic energy is independent of $q$, the effective
potential is a constant.

In a canonical ensemble, the relative probability of finding the
state in element $dpdq$ in phase space is given by $dpdq\exp
\left( -\beta H\right)$. If we are only interested in the
probability of finding the state in $dq$,
we integrate the above over $p$, and obtain%
\begin{eqnarray}
dq\int dp\exp \left( -\beta H\right) &=&dq\exp \left( -\beta U\right) \int
dp\exp \left( -\beta K\right)  \notag \\
&=&dq\left( \frac{2\pi }{\beta }\right) ^{n-1}\exp \left( -\beta \left( U+V_{%
\text{eff}}\right) \right).
\end{eqnarray}%
This is the probability to be used, for example, in the
Monte-Carlo algorithm in the CSAW model \cite{Huang, Huang02}.

\begin{center}
\begin{figure}
        \includegraphics[scale=0.4]{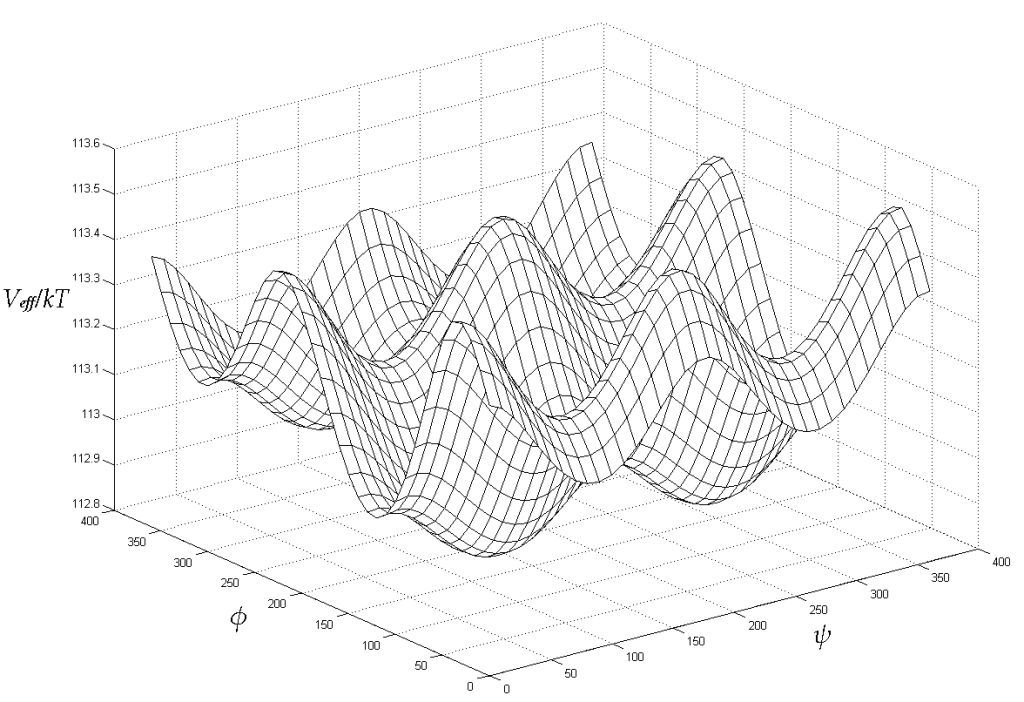}
        \caption{Effective potential of a 3-crank chain at different $\{\phi,\psi\}$ angles between the second and third
        crank. The percentage change of the effective potential is less than
        0.2\%.}\label{fig2}
\end{figure}
\end{center}

We now perform the momentum integration:%
\begin{equation}
\int dp\ e^{-\beta K(p,q)}=\int dp_{1}\cdots dp_{2(n-1)}\exp \left( -\frac{%
\beta }{2}p^{T}M^{-1}p\right) =\left( \frac{2\pi }{\beta }\right) ^{n-1}%
\sqrt{\text{det }M}.
\end{equation}%
Thus%
\begin{equation}
\beta V_{\text{eff}}(q)=-\frac{1}{2}\ln \left( \text{det }M\right)
=-\frac{1}{2}\text{Tr} \left(\ln M \right).
\end{equation}%
Since $V_{\text{eff}}(q)$ depends on all the torsional angles, it
is a function of the chain conformation. Our calculations show
that it is sensibly constant for almost all different
conformations. Representative results are shown in Fig.
\ref{fig2}.

\vspace{-25pt}
\section{Potential energy of hydrogen bonding}

As an application of the canonical formalism, we shall calculate
the normal modes of the $\alpha $-helix and the $\beta $-sheet,
which are important secondary structures in the folded state of a
protein. The main stabilizing agents for these structures are
hydrogen bonds, which exist between N-H and C=O groups from
different residues \cite{KarshikoffBook} \cite{Finkelstein}. We
assume that a hydrogen bond is formed when the distance between
the H and O atoms is $ 2.0\pm 1.0$ \AA , and the bond angle
between N-H and C=O is $180\pm 45^{\circ}$\cite{KarshikoffBook}.

The $\alpha $-helix, also known as the 4$_{13}$-helix, is the most abundant
secondary structure due to its tight conformation \cite{Finkelstein}. In
this configuration, a hydrogen bond connects the C=O group of $i$th crank to
the N-H group of ($i+3$)th crank.

The $\beta $-sheet is a two-dimensional mat made up of backbone
strands stitched together by hydrogen bonds \cite{Finkelstein}.The
participating strands may be parallel or antiparallel.

We wish to study the normal modes of small vibrations about an
equilibrium configuration. The potential energy $U$ is assumed to
be minimum, and taken to be zero, at this configuration. The
equilibrium is assumed to be maintained by hydrogen bonds.
Deviations from equilibrium arise from the stretching and bending
of these bonds. Let $\mathbf{b}_{i}$ be the bond vector of the
$i$th hydrogen bond, i.e. the vector between the O and the bonded
H, in the equilibrium situation. Let $\mathbf{b}_{i}^{\prime }$ be
the same vector when the configuration is displaced from
equilibrium. The
displacement vector is given by%
\begin{equation}
\mathbf{u}_{i}=\mathbf{b}_{i}^{\prime }-\mathbf{b}_{i}.
\label{eq18}
\end{equation}%
For small displacements, we take the potential energy to be
\begin{equation}
U=\frac{1}{2}\kappa _{1}\sum_{i}(|\mathbf{\hat{b}}_{i}\cdot \mathbf{u}%
_{i}|)^{2}+\frac{1}{2}\kappa _{2}\sum_{i}(|\mathbf{\hat{b}}_{i}\times
\mathbf{u}_{i}|)^{2}  \label{eq17}
\end{equation}%
where $\mathbf{\hat{b}}_{i}=$ $\mathbf{b}_{i}/|\mathbf{b}_{i}|$, and $\kappa
_{1}$ and $\kappa _{2}$ are the force constants associated with the
stretching and bending of hydrogen bonds, respectively \cite{Itoh}
\begin{eqnarray}
\kappa _{1} &=&13\text{ N/m },  \notag \\
\kappa _{2} &=&3\text{ N/m }.
\end{eqnarray}

Let the generalized coordinates be denoted%
\begin{equation}
q=q_{0}+\lambda
\end{equation}%
where $q_{0}$ corresponds to equilibrium, and $\lambda $ represents a small
deviation. We can write
\begin{equation}
\mathbf{u}_{i}=\sum_{\alpha }\left( \frac{\partial \mathbf{b}_{i}^{\prime }}{%
\partial q_{\alpha }}\right) _{0}\lambda _{\alpha }+O(\lambda ^{2})
\label{eq19}
\end{equation}%
where the subscript $0$ indicates evaluation at equilibrium. This leads to
the quadratic form
\begin{eqnarray}
U &=&\frac{1}{2}\lambda ^{T}(\kappa _{1}D+\kappa _{2}C)\lambda\ ,
\label{eq21}
\\
D_{\alpha \beta } &=&\sum_{i}\left\vert \mathbf{\hat{b}}_{i}\cdot \frac{%
\partial \mathbf{b}_{i}^{\prime }}{\partial q_{\alpha }}\right\vert
_{0}\cdot \left\vert \mathbf{\hat{b}}_{i}\cdot \frac{\partial \mathbf{b}%
_{i}^{\prime }}{\partial q_{\beta }}\right\vert _{0},  \label{eq22a} \\
C_{\alpha \beta } &=&\sum_{i}\left\vert \mathbf{\hat{b}}_{i}\times \frac{%
\partial \mathbf{b}_{i}^{\prime }}{\partial q_{\alpha }}\right\vert
_{0}\cdot \left\vert \mathbf{\hat{b}}_{i}\times \frac{\partial \mathbf{b}%
_{i}^{\prime }}{\partial q_{\beta }}\right\vert _{0}.  \label{eq22b}
\end{eqnarray}%
\newline
\vspace{-15pt}

\section{Normal Modes}

For small oscillations about equilibrium, the linearized equation of motion
is
\begin{equation}
M\ddot{\lambda}=-\frac{\partial U}{\partial q}\ .  \label{eq24}
\end{equation}%
From (\ref{eq21}) we have
\begin{equation}
\frac{\partial U}{\partial q}=(\kappa _{1}D+\kappa _{2}C)\lambda\
. \label{eq23}
\end{equation}%
Thus
\begin{equation}
M\ddot{\lambda}+(\kappa _{1}D+\kappa _{2}C)\lambda =0.
\label{eq25}
\end{equation}%
The normal frequencies $\omega $ and normal modes $\lambda $ are
eigenvalues and eigenvectors of the equation
\begin{equation}
M^{-1}(\kappa _{1}D+\kappa _{2}C)\lambda =\omega ^{2}\lambda\ .
\label{eq26}
\end{equation}

Our model's validity is subject to the following conditions:

1. We treat small oscillation about a presumed equilibrium configuration $%
q_{0}$. Whether $q_{0}$ indeed corresponds to equilibrium can be verified
through the requirement that all normal frequencies be nonzero and positive.

2. We ignore electrostatic and other interactions. Our results can serve as
a test whether the structures investigated can maintain equilibrium purely
through hydrogen bonding. Inclusion of other interactions will introduce
corrections.

3. Actual $\alpha $ and $\beta $ structures are embedded inside a
protein molecule in solution, and are subject to other forces not
considered here, particularly those arising from Brownian motion
in the solution, the hydrophobic effect, and interaction with
other atoms in the protein. These forces will give rise to
corrections, and may even destroy the stability of the structure.

In view of the limitations of the model, we only examine normal
modes in a frequency range corresponding to wave numbers
$10^{-1}-10^{3}$ cm$^{-1}$. This is because, in a real protein,
the very low-frequency end will be dominated by binding effects to
the rest of the protein, while the very high-frequency region will
be dominated by bond oscillations.

\section{The $\alpha$-helix}

\begin{center}
    \begin{figure}
        \includegraphics[scale=0.5]{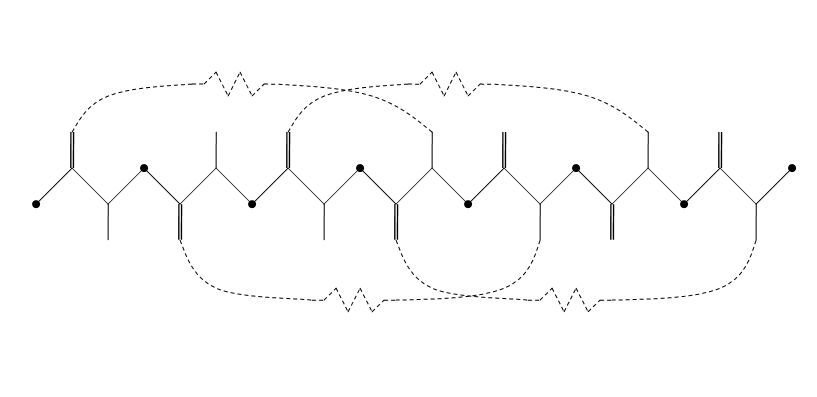}
        \caption{The mechanical system
                    corresponding to small oscillations of the $\protect\alpha
                    $-helix (solid lines). Springs are hydrogen bonds (dashed lines).}\label{fig3}
    \end{figure}
\end{center}

We have modelled a generic $\alpha $-helix using torsional angles $\phi
,\psi $ from polyalanine \cite{KrimmReisdorf}%
\begin{equation}
\{\phi ,\psi \}=\{-57.4^{\circ },\text{ }-47.5^{\circ }\}.
\end{equation}%
The equivalent spring system is illustrated in Fig. \ref{fig3}. In
this example, there are 7 cranks, but only 4 hydrogen bonds. In
general, for $n$ cranks, the number of hydrogen bonds is $n-3.$
The number of degrees of freedom from stretching and bending of
the hydrogen bonds is thus $2(n-3).$ The total number of degrees
of freedom of the system, however, is $2\left( n-1\right) .$ Thus
we expect to have 4 zero modes, apart from rigid translations and
rotations. These will not be included in our results.
\vspace{-30pt}

\begin{center}
\begin{figure}
        \includegraphics[scale=0.5]{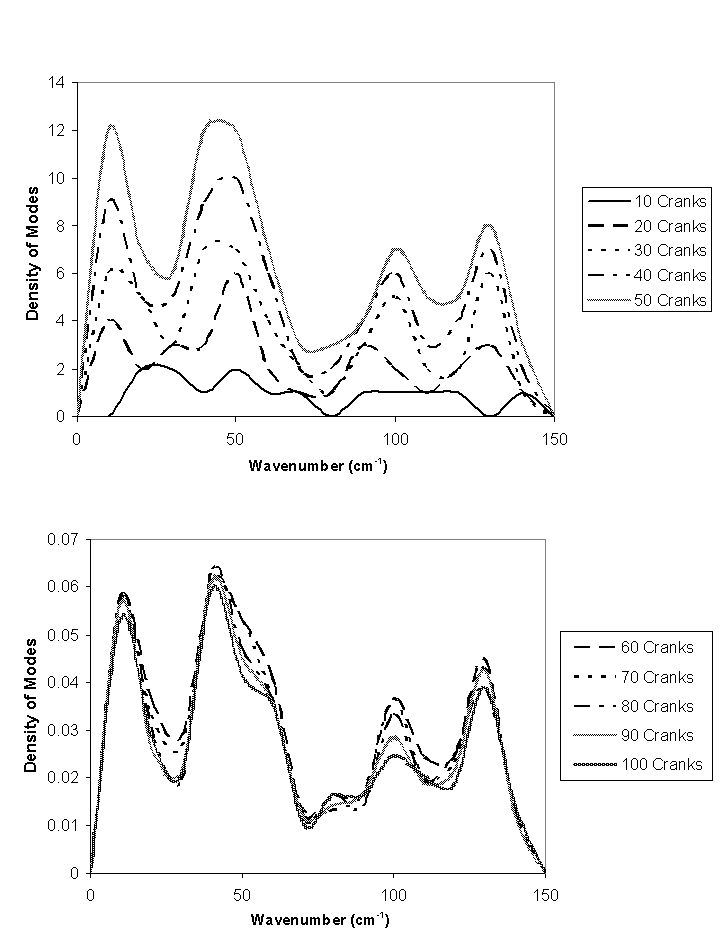}
        \caption{Normal-frequency distributions
                    for the $\protect\alpha $-helix, with different numbers of cranks
                    $n$. Upper panel displays cases $n=10-50.$ Lower panel displays
                    cases $n=60-100$, and the distributions are divided by a scaling
                    factor $n^{1.3}$. Types of distortion corresponding to the peaks
                    are listed in Table \ref{alphavibmotion}.}\label{fig4}
    \end{figure}
    \end{center}

Fig. \ref{fig4} shows the distributions of normal modes as
function of wave number, for different crank numbers $n$. All
calculated frequencies are positive. The upper panel shows
distributions for $n=10-50,$ while the lower panel shows those for
$n=60-100.$ In the latter case, the distributions fit a
scaling law%
\begin{equation}
\text{Density of modes}\varpropto n^{1.3}.  \label{normalize}
\end{equation}%
The plotted distributions have been divided by this factor.
\vspace{-25pt}

\begin{table}[tbp]
\caption{Normal modes of $\protect\alpha $-helix.}
\label{alphavibmotion}%
\begin{center}
\begin{tabular}{|c|c|}
\hline Frequency (cm$^{-1}$) & Mode \\ \hline
$0-10$ & twisting \\
$30-40$ & stretching \& bending \\
$90-100$ & bending \\
$120-130$ & bending \\ \hline
\end{tabular}%
\end{center}
\end{table}

The distributions exhibit 4 peaks associated with various types of
deformation, which can be ascertained by examining the
corresponding eigenvectors. The results are listed in Table
\ref{alphavibmotion}.

\begin{center}
    \begin{figure}
        \includegraphics[scale=0.5]{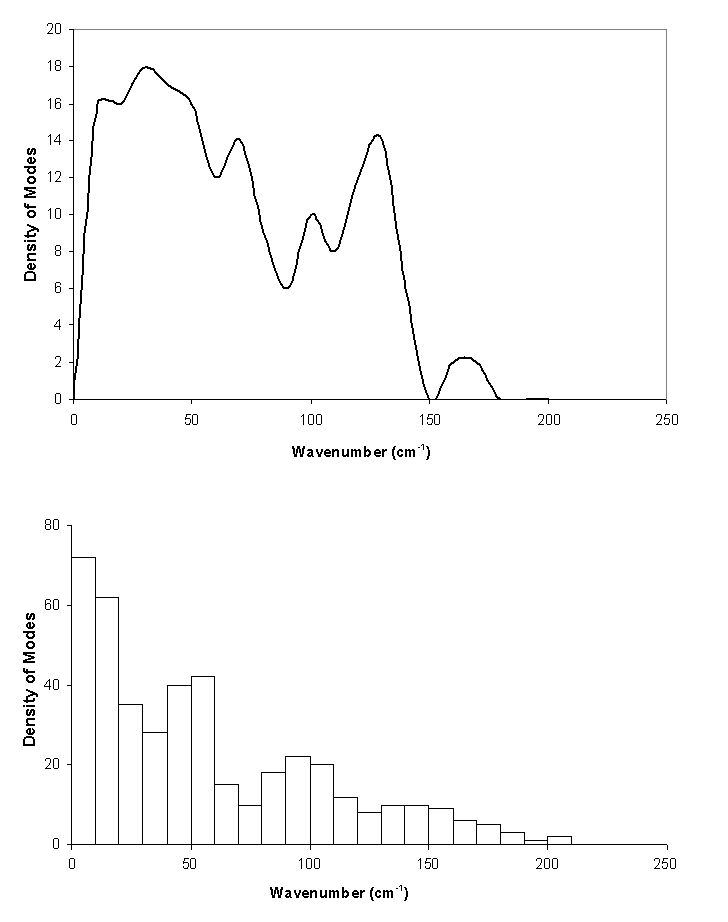}
        \caption{Upper panel: Normal-frequency
                    distribution for myoglobin, constructed by superimposing those of the 8 $%
                    \protect\alpha $-helices in the protein. Lower panel: \ Result of
                    an independent calculation by Krimm and Reisdorf
                    \protect\cite{KrimmReisdorf}. There is qualitative agreement, but
                    the peaks are shifted, possibly due to our neglect of interactions
                    between helices, and contributions from loops.}\label{fig5}
    \end{figure}
\end{center}

As an application of our results, we calculate the normal-mode
distribution for myoglobin (1MBD) \cite{1MBD}, which is made up of
8 alpha-helices, by superposing our calculated distributions. This
procedure ignores the interactions between helices, and
contributions from the loops connecting the helices, and can only
give the crudest approximation to the actual distribution. The
result is shown in the upper panel of Fig. \ref{fig5}. The lower
panel shows a histogram obtained previously by Krimm and Reisdorf
\cite{KrimmReisdorf}, using a different method. There is
qualitative agreement, but the peaks are shifted, presumably due
to interactions neglected in our simple superposition. The rough
shape of the distribution bears resemblance to that calculated for
BPTI, a globular protein with 58 residues \cite{Go}.

\section{The $\beta$-sheet}

\begin{figure}
        \includegraphics[scale=0.5]{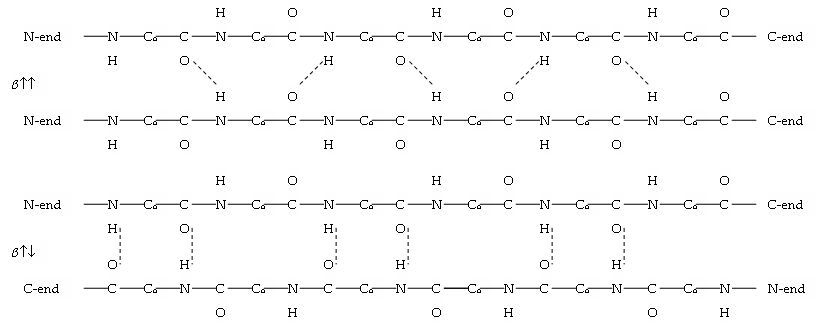}
        \caption{Schematic diagram of $\protect%
                    \beta $-sheets, illustrating the connectivity of hydrogen bonds.
                    Upper diagram: parallel $\protect\beta $-sheet. Lower diagram: antiparallel $%
                    \protect\beta $-sheet.}\label{fig6}
    \end{figure}

We model a generic $\beta $-sheet by setting the torsional angles in each
strand to
\begin{eqnarray}
\{\phi ,\psi \} &=&\{-139^{\circ },\text{ }135^{\circ }\}\text{ \ \
(antiparallel case \cite{MooreKrimm})},  \notag \\
\{\phi ,\psi \} &=&\{-119^{\circ },\text{ }114^{\circ }\}\text{ \
\ (parallel case \cite{BandekarKrimm})}.
\end{eqnarray}%
The connectivity of hydrogen bonds for the parallel and
antiparallel cases is shown in Fig. \ref{fig6}. In the
antiparallel case, an extra crank is included to join two adjacent
strands. In the parallel case, the strands are left open-ended.

Compared to the $\alpha $-helix, the $\beta $-sheet has fewer
hydrogen bonds formed within the structure. Thus we expect that in
our model there will be more zero modes compared to the $\alpha
$-helix; but we ignore them for reasons stated previously.
Otherwise, all calculated frequencies are positive.

\begin{center}
    \begin{figure}
        \includegraphics[scale=0.6]{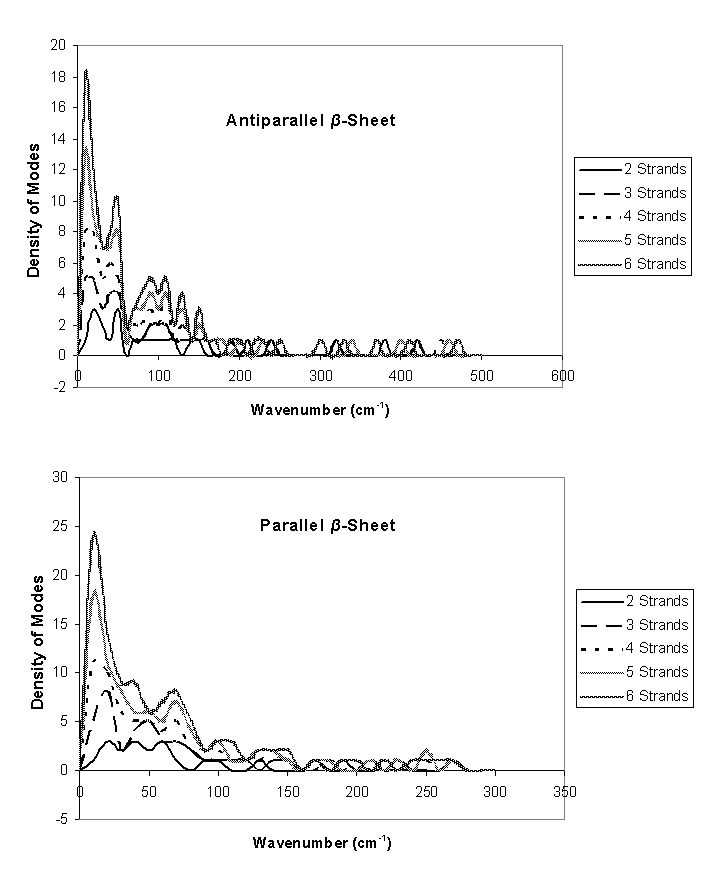}
        \caption{Normal-frequency distributions
                    of $\protect\beta $-sheets, for the same number of cranks per strand, but
                    different number of strands.}\label{fig7}
    \end{figure}
\end{center}

Normal frequencies are computed for varying numbers of strands,
and cranks per strand. We display representative distributions in
Fig. \ref{fig7} for antiparallel and parallel sheets. We see that
the frequencies are concentrated around 50 $cm^{-1}$. This is
consistent with calculations on real protein with $\beta $-sheet
structure \cite{MooreKrimm,AbeKrimm,DwivediKrimm}. In general, the
peak positions of the distributions depend only on the number of
cranks per strand, and are independent of the number of strands.
The peaks tend to widen with increasing crank number.

\section{$\alpha$-$\beta$ transition}

The transition between $\alpha $-helix and $\beta $-sheet is an important
subject, in view of its possible relevance to the prion transition \cite%
{Prusiner}, and the existence of proteins with ambivalent structures \cite%
{Patel}. From our results, we can make a crude calculation, which should be
taken to be of intuitive, rather than practical value.

We can obtain the free energy of a structure in the neighborhood
of equilibrium, by adding the contributions from all of its normal
modes, each treated as a harmonic oscillator. For one classical
harmonic oscillator of natural frequency $\omega ,$ at absolute
temperature $T$, the Helmholtz free energy
is%
\begin{equation}
A_{1}=-k_{B}T\ln \left( \frac{k_{B}T}{\hbar \omega }\right).
\end{equation}%
For $N$ oscillators, corresponding to $N$ normal modes of frequencies $%
\omega _{1},\ldots ,\omega _{N}$, the total free energy is%
\begin{eqnarray}
A_{N} &=&-Nk_{B}T\ln \left( \frac{k_{B}T}{\hbar
\tilde{\omega}}\right);
\notag \\
\tilde{\omega} &=&\left( \omega _{1}\cdots \omega _{N}\right)
^{1/N}.
\end{eqnarray}

\begin{center}
    \begin{figure}
        \includegraphics[scale=0.5]{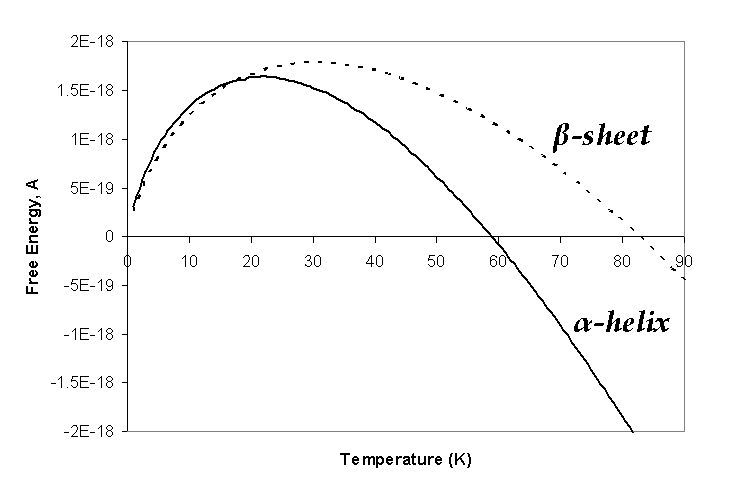}
        \caption{Helmholtz free energy of $%
                    \protect\alpha $-helix and $\protect\beta $-sheet, treated as
                    harmonic oscillators with the corresponding normal frequencies.
                    The intersection of the two curves indicates a phase transition.
                    The equilibrium phase is the one with lower free
                    energy.}\label{fig8}
    \end{figure}
\end{center}

Fig. \ref{fig8} shows the free energies as a function of temperature, for an $\alpha $%
-helix and an antiparallel $\beta $-sheet, each having 39 cranks.
The $\beta $-sheet is made up of 5 strands with 7 cranks per
strand.

The expressions for the free energy are valid only when the system
is harmonic. As models for secondary structures in a real protein,
our results are expected to be valid only in a certain
neighborhood of $k_{B}T/\hbar \tilde{\omega}=1$. \ How large the
neighborhood is depends on interactions between the secondary
structure and its environment. Taking these curves on face value,
we see from Fig. \ref{fig8} that they intersect at $T_{c}=20$ $K$,
above which the $\alpha $-helix has a lower free energy. In this
hypothetical system, then, a transition from $\alpha $-helix to
$\beta $-sheet should occur at $T_{c}$, when the temperature is
lowered.

\bigskip

\newpage

\begin{thebibliography}{99}
\bibitem{Huang} K. Huang, \textit{Biophys. Rev. and Lett}. \textbf{3}, 1
(2008).

\bibitem{Huang02} K. Huang, arXiv: cond-mat/0601244 (2006).

\bibitem{Prusiner} S.B. Prusiner, \textit{Proc. Natl. Acad. Sci. USA},
\textbf{98}, 13363 (1998).

\bibitem{wwpdb}
H. M. Berman, J. Westbrook, Z. Feng, G. Gilliland, T. N. Bhat, H.
Weissig, I. N. Shindyalov and P. E. Bourne, {\it Nucleic Acids
Research}, {\bf 28}, 235 (2002); {\it http://www.pdb.org}.

\bibitem{KarshikoffBook} A. Karshiko, \textit{Non-Covalent Interactions in
Proteins}, (Imperial College Press, London, 2006).

\bibitem{Finkelstein} A.V. Finkelstein and O.B. Ptitsyn, \textit{Protein
Physics: A Course of Lectures}, (Academic Press, London, 2002).

\bibitem{Itoh} K. Itoh and T. Shimanouchi, \textit{Biopolymers}, \textbf{9},
383 (1970).

\bibitem{1MBD}
PDB ID: 1MBD. S. E. Phillips and B.P. Schoenborn, {\it Nature},
{\bf 292}, 81 (1981).

\bibitem{KrimmReisdorf} K. Krimm and W.C. Reisdorf Jr, \textit{Faraday
Discussions}, \textbf{99}, 181 (1994).

\bibitem{Go} N. Go, T. Noguchi, T. Nishikawa, \textit{Proc. Natl. Acad. Sci.
USA}, \textbf{80}, 3696 (1983).

\bibitem{MooreKrimm} W.H. Moore and S. Krimm, \textit{Biopolymers}, \textbf{%
15}, 2465 (1976).

\bibitem{BandekarKrimm} J. Bandekar and S. Krimm, \textit{Biopolymers},
\textbf{27}, 909 (1988).

\bibitem{AbeKrimm} Y. Abe and S. Krimm, {\it Biopolymers}, {\bf
11}, 1817 (1972).

\bibitem{DwivediKrimm} A. M. Dwivedi and S. Krimm, {\it
Macromolecules}, {\bf 15}, 186 (1982).

\bibitem{Patel} S. Patel, P.V. Baleji, and Y.U. Sasidhar, \textit{J. Peptide
Sci}., \textbf{13}, 314 (2007).
\end{thebibliography}
\end{document}